\documentstyle[12pt]{article}
\def\beqa{\begin{eqnarray}}
\def\eeqa{\end{eqnarray}}
\def\beq{\begin{equation}}
\def\eeq{\end{equation}}

\def\half{\frac{1}{2}}
\def\e{\cal E}
\def\l{\cal L}

\def\gd{g_{\mu\nu}}

\def\pa{\partial}

\def\umu{^{\mu}}

\def\dmu{_{\mu}}  
\def\dnu{_{\nu}}

\def\dmunu{_{\mu\nu}}

\def\ie{{\it i.e. }}
\def\eg{{\it e.g. }}
\def\etal{{\it et al.}}

\def\pr{{\it Phys. Rev.}\ }
\def\prl{{\it Phys. Rev. Lett.}\ }
\def\pl{{\it Phys. Lett.}\ }

\def\ijtp{{\it Int. Journ. Theor. Phys.}\ }

\def\apj{{\it Ap. J.}\ }
\def\aa{{\it Astron. Astrophys.}\ }

\def\rmp{{\it Rev. Mod. Phys.}\ }

\begin{document}
\begin{titlepage}

\title{ Oscillating universes as eigensolutions of cosmological
Schr\"odinger equation}

\author{S. Capozziello, A. Feoli, and G. Lambiase$^*$}

	      \date{}
	      \maketitle
\begin{abstract}
We propose a cosmological model which could explain, in a very natural way,
 the apparently periodic
structures of the universe, as revealed in a series of recent observations.
Our point of view is to reduce the cosmological Friedman--Einstein
dynamical system to a sort of Schr\"odinger equation whose bound
eigensolutions are oscillating functions. 
Taking into account the cosmological expansion, the large scale periodic 
structure could be easily recovered considering the amplitudes and the 
correlation lengths of the galaxy clusters.
\end{abstract}

\maketitle

\vspace{20. mm}
\noindent Key--words: cosmology, quantum mechanics, large scale structure, 
general relativity,
Schr\"odinger equation.  

\vspace{20. mm}
\noindent PACS: 98.80-k,98.65.D, 98.80.Hw

\vspace{3. cm}
\noindent $^*${\footnotesize Dipartimento 
di Scienze Fisiche ``E. R. Caianiello'',
Universit\`{a} di Salerno, I-84081 Baronissi (Sa), Italy.
Istituto Nazionale di Fisica Nucleare, Sezione di Napoli, Italy.}\\
{\footnotesize
 e-mail address: capozziello,feoli,lambiase@vaxsa.csied.unisa.it}

\end{titlepage}

\section{\normalsize\bf Introduction}
Understanding the large scale structure of the universe is one of the main
issue of modern cosmology. In fact, a large amount of observations tell us 
that the {\it cosmological principle} breaks  at scales of the order 
$\sim 100 h^{-1}$Mpc and below, while, well above this limit, 
($\sim 1000 h^{-1}$Mpc) it seems that the
matter in the universe is homogeneously and isotropically distributed
(Kolb \& Turner 1990). The symbol $h$ indicates the normalized 
Hubble constant whose values are $0.4\leq h \leq 1$.

From a cosmological point of view (\ie we are considering scales 
where galaxies are pointlike constituents), the clustering properties of 
objects clearly show that a sort of hierarchy exists: we have tight and loose
groups of galaxies, clusters, superclusters and filaments of galaxies. All 
these structures come from an excess of correlation over given distributions 
and they can be well represented by the two--point correlation function
\beq
\label{1}
\xi_{j}(r)=C_{j}r^{-\mu{_j}}\,,
\eeq
which fixes the correlation scales (Kolb \& Turner 1990). The index $j$ can be
$j=G,C,S$ if the correlation is between galaxies, clusters, or superclusters.
$C_{j}$ determines the correlation amplitude which can be from $0.1$
to $10h^{-1}$Mpc for galaxies, up to $25h^{-1}$Mpc for clusters,
and up to $100h^{-1}$Mpc for superclusters (Peebles 1993).
The exponent $\mu_{j}$ assignes the slope of the power law and it is found 
that, for all the hierarchical orders, it is the same, that is 
$\mu_{j}\simeq 1.8$. In other words, the excess of correlation is the 
``same'' for galaxies, clusters and superclusters. This fact could mean that 
the distribution of objects is absolutely non--random and that some fundamental
mechanism has led to the formation of large--scale structure.

A further evidence of such a point of view is the result by 
Broadhurst {\it et al.} (from now on BEKS) (Broadhurst \etal 1990)
which found an excess
of correlation and an apparent regularity in the galaxy distribution with a 
characteristic scale of $128h^{-1}$Mpc. The structure was revealed after the
completion of deep pencil beam surveys extending to red--shift $z>0.2$
toward northern and southern Galactic poles (Broadhurst \etal 1990).
The linear extension of the surveys was to $2000h^{-1}$Mpc so that, the scale
of $128h^{-1}$Mpc between successive density peaks could be interpreted
as a sort of transition scale to homogeneity and isotropy at larger scales.

Several interpretations have been given to this seminal result.
For example, cosmological solutions with an overall Friedmanian expanding 
behaviour, corrected by a small oscillatory regime, could implement such a
structure (Busarello \etal 1994). Furthermore, dynamics of a
nonconformal scalar field could modify cosmological evolution in the sense of
oscillations (Morikawa 1991), (Capozziello \etal 1996).

It emerges that several observational quantities can be affected by such a 
behaviour since oscillations in the cosmological solutions can be easily
reduced to oscillations in the red--shift $z$. In fact, if $a(t)$ is the
cosmological scale factor (function of the cosmic time $t$), we have the
relation
\beq
\label{2}
\frac{\dot{a}}{a}=H=-\frac{\dot{z}}{z+1}\,,
\eeq
so that all quantities containing the Hubble parameter $H$ or the red--shift
$z$ have to oscillate. Then the oscillation in red--shift can be considered,
as first proposed by Tifft (Tifft 1977, Tifft 1987), as a sort 
of ``quantization''.
On the other hand, the apparent periodicity in the distribution of
galaxies implies peaks in the red--shift distribution which lay on
concentric spherical shells with periodically spaced radii
(someone claimed for a new Tolemaic principle from the point of view of 
Earth!).

The main quantities affected by oscillations are the following.
\begin{enumerate}
\item
The number count--red--shift relation of galaxies
\beq
\label{3}
\frac{dN}{d\Omega dL dz}=n(L,t_{0})a_{0}^{2}H^{-1}{\bf d}^{2}\,,
\eeq
where $dN$ is the number of galaxies in the solid angle $d\Omega$ having
red--shift between $z$ and $z+dz$ and luminosity between $L$ and $L+dL$;
 $n(L,t_{0})$ is the number density of galaxies with luminosity $L$ that an
observer sees at $t_{0}$; $a_{0}$ is the actual scale factor and ${\bf d}$
is the comoving distance defined as 
${\displaystyle {\bf d}=\int_{t}^{t{_{0}}}\frac{dt}{a}}$. If $H$ oscillates,
the number of galaxies in a solid angle changes as a function of the 
red--shift and the luminosity.
\item
The two--point correlation function which can be rewritten in term of $H$ as
\beq
\label{4}
\xi(r)=\left(\frac{{\bf d}_{s}}{r_{s}}\right)^{-3}
\left(\frac{{\bf d}}{r}\right)^{2}\left(\frac{H_{0}}{H}\right)-1\,,
\eeq
where $r$ is the red--shift distance (Weinberg 1972)
\beq
\label{5a}
r=zH_{0}^{-1}=\sqrt{\frac{L}{4\pi F}}\,,
\eeq
with $L$ the luminosity of the object and $F$ the measured flux;
$d_{s}$ is the comoving distance of the sample,
$r_{s}$ is the sample size, which can be $r_{s}\simeq 5h^{-1}$Mpc for the 
galaxies and $r_{s}\simeq 25h^{-1}$Mpc for the clusters; $H_{0}$ is the
actual Hubble constant. However, the oscillations can enhance or depress
the galaxy--galaxy correlation function since the distance between the 
objects apparently increases or decreases. 
\item
The number count--redshift relation for quasars. Since the oscillations of 
the Hubble parameter is coherent and global, we expect a similar clustering 
property also in higher red--shift regions such as $1\leq z\leq 6$.
\end{enumerate}
In general, given a sample of homologous objects spaced in red--shift, we 
can, in principle, apply the same argument used for the galaxies, that is we 
can expect an apparent periodicity in the distributions
(\eg Lyman $\alpha$--clouds, radio galaxies, and so on).

Besides we can ask for some fundamental physics explaination 
of such a phenomenology as the above mentioned dynamics of a scalar field.
For example, some authors are recently supposing a deep link between
quantization and gravitation. Some of them support the point of view
that  gravity could give rise to a sort of ``stochastic" quantization
which relates the cosmic scales of interest to the Planck constant
(Calogero 1997). Others search for  extensions of Einstein's relativity
able to include, for examples, fractal structures (Nottale 1997).
A more conservative point of view could be considering cosmic structures
as "remnants" of primordial quantum processes and trying to connect
early ``quantum" dynamics with today--observed classical macroscopic dynamics. 

This is, in some sense, the usual approach
of most of quantum gravity theories which consider either the whole universe
as a quantum system (supposing, for example, several co--existing, 
noninteracting universes (Everett 1957, De Witt 1967, Vilenkin 1982) 
or the universe as a classical
background where primordial quantum processes have given rise to the
actual macroscopic structures (this is the approach of quantum field
theory on curved spacetimes (Birrell \& Davies 1982).

In relation to the first point of view, 
it has been shown that the Wheeler--DeWitt equation ${\cal H}\psi=0$
for the wave function of the universe $\psi$ can be written, in the 
Friedman--Robertson--Walker (FRW) spacetime, as a Schr\"odinger equation
${\displaystyle i\xi\frac{\partial \psi}{\partial \alpha}=\tilde{\cal 
H}}\psi$,
where ${\displaystyle \xi=\frac{d\alpha}{dt}}$ (Pollock 1997).
In this sense (\ie in the mini--superspace approximation), quantum 
gravity can be treated as ordinary quantum
mechanics. This approximation could imply
experimental verifications of the Wheeler--DeWitt equation 
(Pollock 1997, Colella \etal 1975) and give further constraints on big--bang
nucleosynthesis (Llorente \& Per\'ez-Mercader 1995).

In this paper,  following
Rosen 1993, we want to reduce, at least formally, the cosmological 
Einstein-Friedman equations of general relativity to a quantum mechanical
system. If this issue holds, the Friedman equations can be recast as
a Schr\"odinger equation and the 
cosmological solutions can be read as eigensolutions of such a ``cosmological''
Schr\"odinger equation. In this quite simple scheme, the bound states
(with negative energy) are automatically oscillating functions and the 
probability amplitudes can fit the BEKS oscillatory distribution
with an opportune choice of the parameters:
that is, it is nothing else but a particular eigensolution.
This means that the observed oscillations are just the remnant of a 
primordial quantum state enlarged by the cosmological expansion
(which has been huge if an early inflationary phase is supposed).
An important point has to be stressed: here, we are not considering the
above mentioned minisuperspace approach in which one estimates the probability 
that a particular ``classical" universe comes out from quantum boundary
conditions (Everett 1957, De Witt 1967, Vilenkin 1982) : 
we are just recasting the cosmological dynamical 
system as a Schr\"odinger quantum--like system where bounded solutions could 
give rise to the observed large scale structure. The probability has to
be connected to the above correlation function.

The paper is organized as follows. In Sec.2, following the approach by
Rosen (Rosen 1993), we construct our cosmological Schr\"odinger equation.
Sec.3 is devoted to solve such an equation stressing the solutions
characterized by oscillatory behaviours.
Discussion of the results and conclusions are drawn in Sec.4.

\section{\normalsize\bf Cosmological equations as a Schr\"odinger equation}
The cosmological evolution equations come from the Einstein field  equations
\beq
\label{5}
R\dmunu-\half \gd R=\frac{8\pi G}{c^4} T\dmunu+\Lambda\gd\,,
\eeq
where $R\dmunu$ is the Ricci tensor, $R$ is the Ricci scalar, $\gd$ is
the metric tensor, $T\dmunu$ is the stress--energy tensor, and $\Lambda$
is the cosmological constant which is taken into consideration in order to
generalize the discussion. Such equations can be derived from the
Hilbert--Einstein action
\beq
\label{6}
{\cal A}=-\frac{1}{16\pi G}\int d^{4}x\sqrt{-g}\left[R+2\Lambda+
{\cal L}_{m}\right]\,, 
\eeq
where ${\l}_{m}$ is the Lagrangian density relative to the matter fields.
As a simple hypothesis, we consider standard perfect fluid matter so that,
by varying ${\l}_m$ with respect to $\gd$, we get
\beq
\label{7}
T\dmunu=(\rho+p)u\dmu u\dnu-p\gd\,,
\eeq
where $\rho$ and $p$ are the energy--matter density and the pressure
respectively. The dynamical problem is completely 
set when matter evolution
equations are given; they are the contracted Bianchi identities
\beq
\label{8}
T\umu_{\nu;\mu}=0\,.
\eeq
A further equation has to be imposed in order to assign
the thermodynamical state of matter. It, usually, is
\beq
\label{9}
p=(\gamma-1)c^2\rho\;,
\eeq
where $\gamma$ is a constant $(1\leq \gamma \leq 2$ for standard perfect
fluid matter). 

The Einstein--Friedman cosmological equations are  easily recovered
as soon as the Friedman--Robertson--Walker metric 
\beq
\label{10}
ds^2=c^2dt^2-a^2(t)\left\{\frac{dr^2}{1-kr^2}+r^2d\theta^2+
r^2\sin^2\theta d\phi^2\right\}\,,
\eeq
is imposed.
Here $a(t)$ is the scalar factor of the universe, $k=0,\pm 1$ is 
the curvature constant.

Then, from Eqs.(\ref{5}) and (\ref{8}), we obtain the system 
\beq
\label{11}
\frac{\ddot{a}}{a}=-\frac{4\pi G}{3c^2}(\rho+3p)+\frac{\Lambda c^2}{3}\,,
\eeq
\beq
\label{12}
\left(\frac{\dot{a}}{a}\right)^2+\frac{kc^2}{a^2}=
\frac{8\pi G}{3c^2}\rho+\frac{\Lambda c^2}{3}\,,
\eeq
\beq
\label{13}
\dot{\rho}+3\left(\frac{\dot{a}}{a}\right)(\rho+p)=0\,,
\eeq
which has to be completed by Eq.(\ref{9}).

Let us now rewrite the $(0,0)$--Einstein equation (\ref{12}) as
\beq
\label{14}
\half m\dot{a}^2-\frac{m}{6}\left(\frac{8\pi G}{c^2}\rho+\Lambda c^2\right)
a^2=-\half mkc^2\,,
\eeq
where $m$ is a mass which will be specified below.

We can formally write, following Rosen (Rosen 1993)
\beq
\label{15}
T+V=E\,,
\eeq
where
\beq
\label{16}
T=\half m \dot{a}^2\,,
\eeq
\beq
\label{17}
V(a)=-\frac{4\pi m G}{3c^2}\rho a^2-\frac{m\Lambda c^2 a^2}{6}\,,
\eeq
and
\beq
\label{18}
E=-\half mkc^2\,.
\eeq
$T$ has the role of the kinetic energy, $V(a)$ of the potential energy,
and $E$ of the total energy of the system.

Since Eq.(\ref{12}) is the first integral of Eq.(\ref{11}),
the equation of motion of a ``particle" of mass $m$ is recovered, being
\beq
\label{19}
m\ddot{a}=-\frac{dV}{da}\,,
\eeq
and using Eq.(\ref{13}).

By this formalism, we can define a ``particle" momentum 
\beq
\label{20}
\Pi=m\dot{a}\,,
\eeq
and the Hamiltonian
\beq
\label{21}
{\cal H}=\frac{\Pi^2}{2m}+V(a)\,.
\eeq
Using the simple scheme of first quantization, we can substitute
\beq
\label{22}
\Pi\rightarrow -i\hbar\frac{\pa}{\pa a}\,,
\eeq
and then derive the Schr\"odinger equation
\beq
\label{23}
i\hbar\frac{\pa\Psi}{\pa t}=-\frac{\hbar^2}{2m}\frac{\pa^2\Psi}{\pa a^2}+
V(a)\Psi\,,
\eeq
where the wave function is
\beq
\label{24}
\Psi=\Psi(a,t)\,.
\eeq
In our case, $m$ can be interpreted as the mass of a 
galaxy and $|\Psi|^2$ as the
probability to find such a galaxy at a given $a(t)$.
Immediately, we can pass from $a(t)$ to the observable red--shift $z$
by the relation 
\beq
\label{26}
\frac{a_0}{a}=1+z\,,
\eeq
where $a_0$ is the actual scale factor. In this sense, $\Psi=\Psi(z,t)$
is the formal probability amplitude to find a  given object of mass $m$ at
a given red--shift $z$, at time $t$.

Following the ordinary quantum mechanics, a stationary state of energy $E$
is 
\beq
\label{27}
\Psi(a,t)=\psi(a)e^{-iEt/\hbar}\,,
\eeq
and the Schr\"odinger stationary equation is
\beq
\label{28}
-\frac{\hbar^2}{2m}\frac{d^2\psi}{da^2}+V(a)\psi=E\psi\,,
\eeq
or, in terms of $z$, 
\beq
\label{29}
-\frac{\hbar^2}{2m}
\left[\frac{(z+1)^4}{a_{0}^2}\frac{d^2\psi}{dz^2}
+\frac{2(z+1)^3}{a_{0}^2}\frac{d\psi}{dz}\right]+{V}(z)\psi=E\psi\,,
\eeq
being
\beq
\label{30}
{V}(z)=-\left(\frac{4\pi mG}{3c^2}\right)\frac{\rho(z)a_{0}^2}{(1+z)^2}
-\frac{m\Lambda c^2}{6(1+z)^2}\,.
\eeq
We shall continue our analysis by using the scale factor $a(t)$, 
since the  results can be easily translated in $z$.

To conclude this section, we have to stress that our model differs
from that in Rosen ( Rosen 1993), since there $m$ was the mass of the whole
universe, while we are using an effective ``particle", which, from a
cosmological point of view, can be a galaxy. Furthermore, we propose
a different probability interpretation  for $\Psi$.

\section{\normalsize\bf The eigensolutions}
From Eqs.(\ref{9}) and (\ref{13}), we obtain
\beq
\label{32}
\rho=Aa^{-3\gamma}\,,
\eeq
where $A$ is a positive integration constant which gives the density of 
standard matter at a given epoch $a_{0}$. 
The thermodynamical state of matter is
assigned by giving $\gamma$. We have, for example,
 $\gamma=1$ for dust--like matter
and $\gamma=4/3$ for radiation.

The Schr\"odinger stationary equation (\ref{28}) can be written in the form
\beq
\label{33}
\psi''+\left[Ba^{(2-3\gamma)}+Ca^2+{\cal E}\right]\psi=0\,,
\eeq
where
\beq
\label{34}
B=\frac{8\pi G m^2}{3\hbar^2}A\,,\;\;\;\;C=\frac{m^2c^2\Lambda}{3\hbar^2}\,,
\;\;\;\;{\e}=\frac{2mE}{\hbar^2}=-\frac{mc^2k}{2}\,,
\eeq
and the prime indicates the derivative with respect to $a$.
Such a parametric differential equation can be exactly solved in several
cases giving rise to oscillatory behaviours. In any case, being nothing else 
but the one--dimensional problem of ordinary quantum mechanics, we can 
discuss its general features, in particular when oscillatory solutions are 
present. 

\subsection{\normalsize\bf General considerations}

Let us rewrite Eq.(\ref{33}) as
\beq
\label{35}
\psi''+\left[{\cal E}-U(a)\right]\psi=0\,,
\eeq
where
\beq
\label{36}
U(a)=-\left[Ba^{(2-3\gamma)}+Ca^2\right]\,.
\eeq
By comparing the potential $U(a)$ and the ``energy" ${\cal E}$, we can discuss
the qualitative behaviour of solutions. However, the form of  potential
strictly depends on the parameters $B$, $\gamma$, and $C$ (alternatively
on $A$, $\gamma$, and $\Lambda$) so that, due to the meaning which we are 
going to give to $\psi$, the clustering and correlation properties
are functions, as it must be, of the kind and of the content of 
the fluids involved 
into the dynamics.
Furthermore, the energy is "fixed" by the given FRW--spatial model.

If we require that we cannot find galaxies at infinity, 
the function $\psi(a)$ is normalized.
As a trivial result, the function $\psi(a)=0$  is a solution
of the problem. To avoid that this is the {\it only} solution,  we must
ask for nontrivial normalized solutions which will set up a discrete 
spectrum. 

By fixing the energy (that is by choosing the FRW model), we can divide
the $a$--axis in a certain number of intervals using the points where
${\cal E}-U(a)=0$. We assume, for simplicity, that such points
are a finite number. Then, the extreme intervals  have an unbounded extension,
and, between them, we have a finite number of bounded intervals.
We call type--$I$ intervals those where ${\cal E}< U(a)$ so that 
$\psi''/\psi>0$, and type--$II$ intervals those where ${\cal E}>U(a)$, so 
that $\psi''/\psi<0$. In type--$I$ intervals, the function $\psi$ never
changes its sign, in type--$II$ intervals, it is, in general, oscillating
and can change it sign. In this case we could have solutions showing
clustering and anti--clustering properties for galaxies. We have
bending points if ${\cal E}=U(a)$: in these cases, the solution transits
from type--$I$ to type--$II$ intervals and {\it viceversa}.
The function is normalizable if the extreme intervals are of type $I$.
In fact, if we are considering type--$II$ intervals, the ``particle" can
escape towards infinity, that is it can stay at extremely great distance
from a given point. This fact is  physically in contrast with any
clustering or correlation property. In other words, the discrete spectrum
is inside the energy values which do not allow the particle to escape
towards infinity. For this reason, the energy must be less than
$U(a)$.
However, we have to stress that 
the situation $t\rightarrow -\infty$ means that we are considering
singularity--free models but $a(t)$ has to remain nonnegative.

 If we are in some regime where $|{\cal E}|\gg U(a)$, with
${\cal E}$ positive defined (\ie $k=-1$), Eq.(\ref{35}) has stationary
wave solutions of the form
\beq
\label{37}
\psi(a)=a_1 e^{i\sqrt{\cal E}a}+a_2 e^{-i\sqrt{\cal E}a}\,,
\eeq
if the energy is negative, the physical solutions are decreasing exponential 
functions.
Such functions are not always normalizable but it is easy to see that
oscillatory behaviours can be recovered.

\subsection{\normalsize\bf Special cases}
From Eq.(\ref{35}), by choosing suitable values of the parameters
$B$, $\gamma$, $C$, and ${\cal E}$, we can recover all the standard
cases of ordinary quantum mechanichs (potential well, harmonic oscillator,
Coulomb--like potential, and so on). As it is well known (Messiah 1961),
many of them exhibit oscillatory eigensolutions.
Considering our point of view, the interesting values of $\gamma$
are $\gamma=4/3$ (radiation), $\gamma=1$ (dust), $\gamma=0$
(scalar field).
\subsubsection{\normalsize \it The radiation case}
In this case, Eq.(\ref{33}) becomes of the form
\beq
\label{38}
\psi''+\left[\frac{B}{a^{2}}+Ca^2+{\cal E}\right]\psi=0\,.
\eeq
For $\Lambda=0$ and $k=-1$, it can be written as
\beq
\label{39}
\psi''+\left[\beta^2-\frac{(\nu^2-1/4)}{a^2}\right]\psi=0\,,
\eeq
with 
\beq
\label{40}
\beta^2={\e}=\frac{mc^2}{2}\,,\;\;\;\;
\nu=\sqrt{\frac{1}{4}-B}\,,
\eeq
which is  a Bessel equation.
The general solution is (Abramowitz \& Stegun 1970)
\beq
\label{41}
\psi(a)=\sqrt{a}Z_{\nu}\left(\beta a\right)\,.
\eeq
From the theory of Bessel functions, $Z_{\nu}(x)$ can be given as a 
combination
of the functions $J_{\nu}(x)$, $Y_{\nu}(x)$, or $H_{\nu}^{(1)}(x)$, 
$H_{\nu}^{(2)}(x)$. The analysis of asymptotic behaviours results 
particularly interesting. We have, for $a\rightarrow 0$, 
\beqa
\label{42}
J_{\nu}(a)&\sim&\left(\frac{a}{2}\right)^{\nu}\Gamma(\nu+1)\,,
\nonumber\\
Y_{0}(a)&\sim&-iH_{0}^{(1)}(a)\sim iH_{0}^{(2)}(a)\sim\frac{2}{\pi}\ln a\,,
\nonumber\\
Y_{\nu}(a)&\sim&-iH_{\nu}^{(1)}(a)\sim iH_{\nu}^{(2)}(a)\sim
-\frac{1}{\pi}\Gamma(\nu)\left(\frac{a}{2}\right)^{-\nu}\,.
\eeqa
In the opposite situation, that is $a\rightarrow\infty$, we have oscillating
solutions
\beqa
\label{43}
J_{\nu}(a)&\sim&\sqrt{\frac{2}{\pi a}}\cos\left(a-\frac{\nu\pi}{2}-
\frac{\pi}{4}\right)\,,
\nonumber\\
Y_{\nu}(a)&\sim&\sqrt{\frac{2}{\pi a}}\sin\left(a-\frac{\nu\pi}{2}-
\frac{\pi}{4}\right)\,,
\nonumber\\
H_{\nu}^{(1)}(a)&\sim&\sqrt{\frac{2}{\pi a}}\exp\left[i\left(
a-\frac{\nu\pi}{2}-\frac{\pi}{4}\right)\right]\,,\\
H_{\nu}^{(2)}(a)&\sim&\sqrt{\frac{2}{\pi a}}\exp\left[-i\left(
a-\frac{\nu\pi}{2}-\frac{\pi}{4}\right)\right]\,.
\eeqa
For $\Lambda=0$ and $k=+1$, we get
\beq
\label{44}
\psi(a)=\sqrt{a} Z_{\nu}\left(i\sqrt{|{\e}|}a\right)\,,
\eeq
where the relations
\beqa
\label{45}
J_{\nu}(ia)&=&e^{i\pi\nu}I_{\nu}(a)\,,\nonumber\\
Y_{\nu}(ia)&=&e^{i(\nu+1)\pi/2}I_{\nu}(a)-
\frac{2}{\pi}e^{-i\pi\nu}K_{\nu}(a)\,,\nonumber\\
H_{\nu}^{(1)}(ia)&=&-i\frac{2}{\pi}e^{-i\pi\nu/2}K_{\nu}(a)\,,
\eeqa
hold.

If $k=0$ but $\Lambda\neq 0$, we have
\beq
\label{46}
\psi(a)=\sqrt{a}Z_{\nu}\left(\frac{\sqrt{C}}{2}a^2\right)\,,
\eeq
where 
\beq
\label{47}
\nu=\frac{\sqrt{1-4B}}{4}\,.
\eeq

\subsubsection{\normalsize \it The dust case}

In the dust--dominated case, we have
\beq
\label{48}
\psi''+\left[\frac{B}{a}+Ca^2+{\cal E}\right]\psi=0\,,
\eeq
and for $k=\Lambda=0$, we have solutions of the form
\beq
\label{49}
\psi(a)=\sqrt{a}Z_{\nu}(\sqrt{a})\,,
\eeq
where the above considerations hold.
If $C=0$, we get an hydrogenlike Schr\"odinger equation (with null angular 
momentum) and the wavefunction $\psi(a)$ can be written as a combination 
of Laguerre polynomials (see also Rosen 1993, Messiah 1961). 

\subsubsection{\normalsize \it The scalar field case}
The scalar field case, is not exactly a standard perfect fluid form
of matter but it is of extreme physical interest since it is connected
to inflation (Kolb \& Turner 1990). However, it can be treated as 
a standard fluid
assuming $\gamma=0$ in the equation of state Eq.(\ref{9}) (Guth 1981).
From the point of view of dynamics, it has the same role of cosmological
constant so that Eq.(\ref{33}) becomes
\beq
\label{50}
\psi''+\left[Ca^2+{\cal E}\right]\psi=0\,.
\eeq
We get the same equation by putting $B=0$. However, the cosmological constant
must be redefined. It is clear that, depending on the relative signs and 
values of $C$ and ${\cal E}$, Eq.(\ref{50}) reduces to the equation of
an harmonic oscillator with combinations of Hermite polynomials as
eigensolutions or, in general, to the differential equation of
parabolic cylinder functions (Abramowitz \& Stegun 1970).

\section{\normalsize\bf Discussion and Conclusions}

Several theoretical interpretations of BEKS's 
result have been attempted so far
 (Busarello \etal 1994, Morikawa 1991,
Capozziello 1996, Budinich \& Raczka 1993) but in
all of them   modifications of Einstein standard theory of
gravity have been invoked. Essentially, a further ingredient as a scalar
field should have had the role of perturbing the overall Friedmanian
dynamics of the universe breaking the homogeneity and isotropy at
scales of the order of $\sim 100h^{-1}$Mpc.
The approach which we have adopted in this paper is, in some sense,
absolutely conservative since we are still using Einstein general relativity.
The only issue is that the Friedman--Einstein cosmological dynamics
could be read in the sense of elementary quantum mechanics: the
$(0,0)$--energy equation should give the Schr\"odinger equation in a simple
scheme of first quantization. In order to justify such an assumption,
we must suppose a primordial quantum dynamics of matter contained in the
universe and the fact that the today observed large scale structure reflects
the phenomenology of early epochs. In other words, we can say that
the Schr\"odinger Eq.(\ref{33}) describes exactly the behaviour of the universe
for any value of the cosmic time and the classical Einstein equations
(in particular the $(0,0)$)
well reflect the evolution of dynamical system when it can be described
at a classical level.

Here we have not considered any questions
connected neither to quantum field theory nor to quantum gravity.
We are simply  taking into account the ``particle" dynamics of objects which
are galaxies. The probability of finding them at a certain epoch (\ie $a(t)$)
or a certain red--shift is given by the Schr\"odinger equation which
is nothing else but the quantum counterpart of the classical Einstein
first order equation. Furthermore, the eigensolutions of such an equation,
having a probabilistic meaning, can be connected to the correlation function
which gives the features of the distribution of a certain class of objects
(\eg galaxies, clusters or superclusters of galaxies). 

As we have shown, several possibilities exist to get  oscillatory behaviours:
they depend on the kinds and the number of fluids into the dynamics and,
essentially, on the kind of FRW spatial model. If the universe is
dust--dominated, with a remnant of cosmological constant, the asymptotic
behaviour of a Bessel function could easily reproduce the BEKS observations
(in Budinich 1993, Gegenbauer's polynomial are used).
A good match is obtained if $\psi(a)$, that is $\psi(z)$, has 8 oscillations
with periodicity of $128h^{-1}$Mpc in $2000h^{-1}$Mpc (\ie in a red--shift
range from $z=0$ to $z=0.5$). The amplitude of such oscillations strictly
depends on the cosmological densities of fluids involved (see Busarello
\etal 1994 for a detailed discussion).

In conclusion, we can say that a breaking of homogeneity and isotropy
at small scales ($\sim 100h^{-1}$Mpc) with oscillating correlations
(and anticorrelations) between galaxies could be easily implemented 
considering a sort of cosmological Schr\"odinger equation.

\begin{center}
{\bf References}
\end{center}

Abramowitz M. and I. Stegun, {\it Handbook of Mathematical Functions}
Dover (1970) New York.\\
Birrell N.D.  and P.C.W. Davies, {\it Quantum fields in curved space}
Cambridge Univ. Press (1982) Cambridge. \\
Broadhurst T.J. \etal {\it Nature} {\bf 343} (1990) 726.\\
Budinich P. and R. Raczka, {\it Found. of Phys.} {\bf 23} (225) 1993.\\
Busarello G. {\it et al.}, \aa {\bf 283} (1994) 717.\\
Calogero F. \pl {\bf 228 A} (1997) 335.\\
Capozziello S. \etal {\it La Riv. del Nuovo Cim.} {\bf 4} (1996) 1.\\
Colella R. \etal \prl {\bf 34} (1975) 1472.\\
De Witt B.S. \pr {\bf 160} (1967) 1113.\\
Everett H. \rmp {\bf 29} (1957) 454.\\
Guth A. \pr {\bf 23 D} (1981) 347.\\
Kolb E.W. and M.S. Turner {\it The Early Universe} (1990)
Addison--Wesley, New York.\\
Llorente A. and J. Per\'ez--Mercader, \pl {\bf 355 B} (1995) 461.\\
Messiah A. {\it Quantum Mechanics} North--Holland (1961) Amsterdam.\\
Morikawa M. \apj {\bf 369} (1991) 20.\\
Nottale L. \aa {\bf 327} (1997) 867.\\
Peebles P.J.E. {\it Principles of Physical Cosmology} (1993)
Princeton Univ. Press, Princeton.\\
Pollock M.D. {\it Mod. Phys. Lett.} {\bf 12 A} (1997) 2057.\\
Rosen N. \ijtp {\bf 32} (1993) 1435.\\
Tifft W.G. \apj {\bf 211} (1977) 31.\\
Tifft W.G. in {\it New Ideas in Astronomy} (1987) eds. F. Bertola.\\
Vilenkin A. \pl {\bf 117 B} (1982) 25.\\
Weinberg S. {\it Gravitation and Cosmology} (1972) ed. Wiley, New York.

\vfill
\end{document}